\documentclass[amstex,12pt]{article}     
\usepackage{amssymb,amsmath}           

\newlength{\dinwidth}
\newlength{\dinmargin}
\newtheorem{see}{}[section]

\newcommand{\ie}{{\it i.e.\ }}
\newcommand{\eg}{{\it e.g.\ }}
\newcommand{\cf}{{\it cf.\ }}

\newcommand{\nind}{\noindent}
\newcommand{\bmu}{{\mbox{\boldmath $\mu$}}}
\newcommand{\sbmu}{{\mbox{\footnotesize \boldmath $\mu$}}}

\newcommand{\RR}{\mathbb R}
\newcommand{\CC}{\mathbb C}
\newcommand{\NN}{\mathbb N}
\newcommand{\ZZ}{\mathbb Z}

\newcommand{\supp}{\mbox{supp}}

\newcommand{\Cb}{{\cal C}_\beta}
\newcommand{\CLb}{{\cal C}_{\Lambda  \beta}}
\newcommand{\CD}{{\cal C}_\Delt}
\newcommand{\CU}{{\cal C}}

\newcommand{\bp}{{\mbox{\boldmath $p$}}}
\newcommand{\bx}{{\mbox{\boldmath $x$}}}

\newcommand{\Delt}{{B}}
\newcommand{\EE}{\mbox{E}}
\newcommand{\TF}{F}

\begin{document}
\title {Thermodynamic Properties of Non-Equilibrium States in 
Quantum Field Theory}
\author{Detlev Buchholz\,$^a$,  Izumi Ojima\,$^b$  and  Hansj\"org 
Roos\,$^a$\, \\[5mm]
${}^a$ Institut f\"ur Theoretische Physik, Universit\"at G\"ottingen\\
37073 G\"ottingen, Germany\\[2mm]
${}^b$ Research Institute for Mathematical Sciences, Kyoto
University\\ Kyoto 606--8502, Japan}
\date{}
\maketitle 
\begin{abstract}{\noindent Within the framework of 
relativistic quantum field theory, a novel method is established
which allows to distinguish non-equilibrium   
states admitting locally a thermodynamic interpretation. 
The basic idea is to compare 
these states with global equilibrium states (KMS states) by means of 
local thermal observables. With the help of such observables, the 
states can be ordered into classes of increasing local thermal
stability. Moreover, it is possible to identify states 
exhibiting certain specific thermal properties of interest, 
such as a definite local temperature or entropy density. 
The method is illustrated in a simple model describing
the spatio-temporal   evolution of a ``big heat bang''.} 
\end{abstract}
\section{Introduction} 
\setcounter{equation}{0}

States of macroscopic systems which are in global thermal 
equilibrium can be distinguished in a clearcut manner 
within the setting of quantum statistical mechanics 
by the Kubo-Martin-Schwinger (KMS) condition \cite{HaHuWi}.
This condition, which
comprises the pertinent features of the Gibbs ensemble
in the case of finite and infinite quantum systems 
\cite{BrRo}, has proved to be a powerful
ingredient, both, in the construction of equilibrium 
states in concrete models and in their general structural 
analysis. 

However, there does not exist  a similarly general 
and powerful characterization of states complying  
with the heuristic idea of being only locally in 
thermal equilibrium, \eg having only locally a definite 
temperature which varies from point to point. 
The physical origin of this conceptual difficulty 
is obvious: if one deviates from a global equilibrium
situation, there appear a variety of possibilities, 
ranging from mild perturbations of equilibrium 
states through steady states, whose properties 
are governed by external heat baths, or hydrodynamic flows  
up to systems which no longer 
admit a description in terms of thermodynamic notions. Thus,
what is required are concepts allowing to distinguish
between these different situations and describing  the
respective local thermal properties of the corresponding classes of 
states. 

There exists an extensive literature on this subject,
\cf the comprehensive treatment in \cite{ZuMoRo} and
references quoted there.  Most approaches are based 
on a maximal entropy principle, suggesting a modified 
Gibbs ansatz in a suitable (large time) limit of the theory.  Other
methods  proceed by some procedure of ``coarse graining'' directly  
to an effective macroscopic description.
The conceptual basis of these approaches, however,
does not always seem  clear. In any case, the question 
of identifying  the states 
of interest in the microscopic theory is left aside. 

In the present article we analyse systematically  
this problem in the setting of 
relativistic quantum field theory. This framework  
is particularly well suited for such an investigation 
since it provides an assignment of observables to 
space-time regions, yielding a description of the local
properties of physical states. In particular, it allows
a local comparison of different states,      
which has proved to be an important element  
of structural analysis \cite{Ha}. Within the present 
thermodynamic context, it was used as an 
ingredient in discussions of local thermal aspects 
of physical states in \cite{BuJu,Na,He} and we will 
also make use of it here. 

Before going into technical details, let us outline
the heuristic ideas underlying our approach. 
The basic ingredients are suitable ``local thermal 
observables'', assigned to the space-time points $x$,
by which any given state can be compared with the 
members of the family of global equilibrium states. If a state happens 
to coincide at $x$ with a  global equilibrium state 
to some degree of accuracy, \ie if the 
expectation values of a sufficiently large number of 
thermal observables at $x$ coincide in the two states,   
it is meaningful to say that the given state is 
approximately in equilibrium at $x$. The larger the 
number of observables for which the two expectation 
values agree, the better the state complies with 
the idea of local equilibrium.
 
The  use of local observables for the description
of thermal properties requires some justification 
since thermodynamic notions are usually regarded 
as being of a macroscopic nature. Let us discuss this 
for the case of temperature. What one needs to measure temperature
is a suitable thermometer, \ie a device which allows discrimination of
equilibrium states of different temperatures. In dealing with 
very large systems in global equilibrium it does not matter 
where and when this measurement takes place. The thermometer may 
be coupled to any part of the system under consideration 
at any time. If one increases  its sensitivity, 
its spatial extension can be taken 
to be arbitrarily small. Moreover,  
as the system is in a stationary state, time averages  
can be replaced by the mean value of sufficiently many 
measurements at arbitrarily short time intervals.
Therefore, for a system in global equilibrium, it is
possible to determine the temperature 
by observables in arbitrarily small space-time
cells, provided the corresponding measurements are repeated
sufficiently often. The smaller the cell the larger are the
fluctuations of the results of the individual measurements; but the
collection of these data provides the desired information.  

This simple observation suggests the 
following definition of local temperature in a given  
space-time cell of a non-equilibrium system (provided a whole
ensemble, \ie a large number of copies of the system, can be prepared):  
take sufficiently many readings of a suitable observable in that
particular cell and compare them with those in global equilibrium
states of arbitrary temperature. If the two sets of data happen to
coincide for some value of the temperature, it is meaningful to
ascribe that temperature value to the chosen cell of the
non-equilibrium system under consideration. Thus the local temperature
is to be regarded as a property of the ensemble, not of an
individual member of it. But as quantum theory is a statistical
theory, this point of view fits naturally into its mathematical
framework.

It is conceptually convenient to proceed from the measurements in 
finite space-time cells to the idealization of temperature measurements 
performed at a space-time point. Such measurments can, however, no 
longer be described by observables 
(operators); one has to use quadratic forms. But since this causes no 
major mathematical difficulties, it seems natural to 
perform this additional step.   

The idea of using local observables (or quadratic forms)
for the description of the thermal properties
of non-equilibrium states can be extended to other intensive  
thermal parameters of interest such as the energy density
and the charge density. Again one has to exhibit suitable local
observables which are sufficiently sensitive 
to these properties and to calibrate them by using the 
global equilibrium states as reference systems. The measured 
data taken in a non-equilibrium state have to be  
compared with those recorded in global equilibrium 
and if these data agree for a particular reference    
state one can attribute the values of the corresponding thermal 
parameters to the state describing the non-equilibrium 
situation. If, in particular, a state admits locally 
an interpretation in this manner  
in terms of a unique global equilibrium state, 
then all its local thermal parameters are fixed, including fundamental 
quantities such as the entropy density. 

Transforming this idea into action, we have to cope with a difficulty
which we have ignored so far: the notion of equilibrium in general 
refers to the state of a system at rest with respect to the observer. 
Dealing with an arbitrary state (describing, 
{\it e.g.}, a hydrodynamic flow) 
we have to take into account that these rest systems may 
vary from point to point. Consequently, the set of 
thermal observables should be sufficiently rich to determine
these local rest systems and the set of thermal reference states
should include all Lorentz-transformed equilibrium states, too.
Moreover, in order to cover also situations, where the thermal 
parameters do not have sharp values, we have also to admit mixtures of
global equilibrium states as thermal reference states.   

Thus, to summarize, a properly chosen set of local thermal observables 
together with the set of thermal reference states, fixed by the theory, 
will be our tool for the analysis and interpretation 
of the local thermal properties of non-equilibrium states. 

After this exposition of the heuristic ideas underlying our approach,
let us turn now to its mathematical formulation. In the next 
section we recall some basic elements of thermal quantum 
field theory and introduce the thermal 
reference states. Section 3 contains the characterization 
of the local thermal observables. In  
Sec.\ 4  we identify with their help  
a hierarchy of physical states which may be regarded 
as being locally close to thermal equilibrium with increasing 
degree of accuracy. An instructive model, illustrating 
various points of our general analysis,    
is disussed in Sec.\ 5 and the article concludes with 
remarks on some conceptual aspects of our approach. In the  
appendix we establish a theory of
thermostatics for the set of thermal reference states used in the main text.
 
\section{Thermal reference states}
\setcounter{equation}{0}

As we are dealing here with a conceptual problem,  
the present investigation is carried out in the general
setting of thermal quantum field theory. We begin by recalling  
some relevant notions and results, primarily in order to fix our
notation. 

The basic objects in our analysis are local observable 
fields, such as conserved currents, the stress energy tensor, etc., 
denoted generically by $\phi (x)$ (with tensor indices omitted). 
Quantum fields at a point are known to be quite   
singular objects which are only defined in the sense of quadratic
forms on a suitable domain, \cf the subsequent section.
It is therefore common to proceed from the point fields $\phi(x)$ to 
averages with suitable test functions $f$ having compact 
support in Minkowski space $\RR^4$,
\begin{equation} \label{e.smear}
\phi (f) \doteq \int \! dx \, f(x) \phi (x). 
\end{equation}
These are defined as operator-valued distributions on some common
dense and  
stable domain $\cal{D}$ in the 
underlying Hilbert space $\cal{H}$, where the theory has been
constructed (usually the vacuum sector).
The finite sums and products of the averaged field operators 
generate a *-algebra $\cal{A}$, called the algebra of observables, 
consisting of all polynomials $A$ of the form  
\begin{equation} \label{polynom}
A=\sum \phi (f_{1}) \phi (f_{2})\cdots \phi (f_{n}). 
\end{equation}
We assume that also all multiples of the unit operator $1$ are 
elements of ${\cal A}$. As the fields $\phi$ are real 
(being observable), the  
*-operation (Hermitian conjugation) on $\cal{A}$ is defined by 
\begin{equation} \label{involution}
A^{\ast } \doteq \sum \phi(\overline{f_{n}}) \cdots 
\phi (\overline{f_{2}}) \phi (\overline{f_{1}}),
\end{equation}
where $\overline{f}$ denotes the complex conjugate of $f$. 
The Poincar\'e 
transformations $\lambda \in {\cal P}_+^\uparrow$ act on $\cal{A}$
by automorphisms $\alpha_\lambda$,  
\begin{equation} \label{e.transl}
\alpha_\lambda (A) \doteq
\sum \phi (f_{1,\lambda}) \phi (f_{2,\lambda})\cdots \phi (f_{n,\lambda}), 
\end{equation}
where $f_\lambda$ is defined by 
$f_\lambda (x) \doteq D(\lambda ) f(\lambda^{-1} x)$ 
and $D$ is a matrix
representation of ${\cal P}_+^\uparrow$ corresponding to the tensor
character of the respective field $\phi$. If $\lambda$ is a pure
translation, $\lambda = (1,a)$, $a \in \RR^4$, 
we denote the corresponding automorphism
by $\alpha_a$. Note that 
the action of these automorphisms on $\cal{A}$ is well-defined; there
is no need  to assume that they are unitarily implemented. 

As we restrict attention here to observable fields, there is one
further important ingredient in this general setting, namely 
locality (Einstein causality). It is the requirement that fields
localized at spacelike separated points commute. We will make
use of this property at some technical points in the subsequent analysis.
  
The states of the physical systems are described by 
positive normalized expectation
functionals on the algebra $\cal{A}$, generically denoted by $\omega$. 
We recall that these functionals have the defining properties 
$\omega (c_{1}A_{1}+c_{2}A_{2}) = c_{1} \, \omega(A_{1}) +c_{2} \, 
\omega(A_{2}) $, $c_1,c_2 \in \CC$  (linearity),  
$\omega(A^*A) \geq 0$ (positivity), and  
$\omega(1) =1$ (normalization). Basic examples are states 
of the form 
\begin{equation} \label{normal}
\omega (A) = \mbox{Tr} \, \rho A, \quad A \in {\cal A},
\end{equation}
where $\rho$ is a density matrix in the underlying Hilbert space
${\cal H}$
and $\mbox{Tr}$ denotes the corresponding trace. It is, however, 
important to note that not all states of interest here can 
be represented in this way. For, if ${\cal H}$ is the vacuum
Hilbert space, all states of the form (\ref{normal})
describe mixtures of local excitations of the vacuum;
thus global thermal states are not among them.  
An explicit formula for the thermal states, 
analogous to (\ref{normal}), would require a change of 
the Hilbert space representation of the algebra of observables 
${\cal A}$. Since we deal here with states describing very different 
physical ensembles, it is, however, more convenient 
to rely on the abstract characterization of states given 
above. We mention as an aside that one can recover from 
any such state an explicit representation 
in terms of density matrices on some Hilbert space 
by the Gelfand-Naimark-Segal reconstruction theorem \cite{Ha}.
But we make no use of this fact here. 

As has been outlined in the introduction, an important 
ingredient in the present investigation 
are the global equilibrium states which are characterized by 
the KMS condition \cite{BrRo}. Given a Lorentz frame, fixed by 
some positive time-like vector $e \in V_+$ of unit length, this 
condition can be stated as follows.

\nind {\bf KMS condition:} A state $\omega_\beta$ satisfies 
the KMS condition at inverse temperature $\beta > 0$ in the 
given Lorentz frame if for each pair of operators
$A,A^\prime \in {\cal A}$ there is some function $h$ which is analytic  
in the strip $S_\beta \doteq \{ z \in \CC : 0 <
\mbox{Im} z < \beta \} $ 
and continuous at the boundaries such that 
\begin{equation}
h(t) = \omega_\beta(A^\prime \alpha_{te}(A)), 
\ \ h(t+i\beta) = \omega_\beta(\alpha_{te}(A) A^\prime), 
\quad t \in \RR.
\end{equation}
In this situation $\omega_\beta$ is called a KMS state. \\[2mm]
\indent We will have occasion to make also use of a slightly stronger version
of the KMS condition, proposed in \cite{BrBu}. It can be established
under quite natural constraints on the number of local degrees of
freedom of the underlying theory and is a remnant of 
the relativistic spectrum condition in the vacuum sector. It was 
therefore called relativistic KMS condition in \cite{BrBu}. 

\nind {\bf Relativistic KMS condition:} A state $\omega_\beta$ is said 
to satisfy the relativistic KMS condition at inverse temperature 
$\beta > 0$ in the given Lorentz frame  
if for each pair of operators
$A,A^\prime \in {\cal A}$ there is some function $h$ which is analytic in the 
tube $\RR^4 + i \,\big( V_+\cap(\beta e -  V_+) \big)$ such that in the 
sense of continuous boundary values \cite{BrBu} 
\begin{equation}
h(a) = \omega_\beta(A^\prime \alpha_a(A)), \ \ 
h(a + i \beta e) = \omega_\beta(\alpha_a(A) A^\prime), \quad a \in \RR^4.
\end{equation}

It is a well established fact that any KMS state 
$\omega_\beta$ describes an ensemble which is in thermal equilibrium
in the distinguished Lorentz frame, describing the rest system 
of the state. As we have to keep track of both,  temperatures 
and rest systems, it is convenient
to combine this information into a four vector $\beta e \in V_+$,
which will again be denoted by $\beta$. The 
temperature and the rest system 
can be recovered from any such 
temperature vector $\beta \in V_+$ in an obvious manner.

For given $\beta \in V_+$, the corresponding convex set $\Cb$
of all KMS states $\omega_\beta$ is known to form a simplex 
under quite general conditions \cite{BrRo}. This implies in particular
that all states in $\Cb$ can be distinguished with the help
of ``classical'' (central) observables.
Familiar examples of such observables are the chemical potential 
in theories with a conserved charge or the mean density distinguishing
different phases. In order to simplify the subsequent 
discussion, we assume here that each family $\Cb$ 
is non-degenerate, \ie for any given  $\beta$ there is a unique 
KMS state $\omega_\beta$. The case of more complex 
families $\Cb$ can be treated similarly 
but requires some heavier notation.
 
Let us briefly discuss the consequences of this uniqueness assumption 
for the transformation properties of the KMS states under Poincar\'e 
transformations. For any given  
KMS state $\omega_\beta$ and transformation $\lambda \! = \! (\Lambda,a)$,
the corresponding Poincar\'e-transformed 
state is given by  $\omega_\beta \circ \alpha_\lambda^{-1}$,
where the circle denotes the composition of maps. It follows 
from the definition of KMS states that 
$\omega_\beta \circ \alpha_\lambda^{-1}$ is again a KMS state belonging 
to the family $\CLb$. In view of our uniqueness assumption 
we thus obtain the simple transformation formula
\begin{equation} \label{transformation}
\omega_\beta \circ \alpha_\lambda^{-1} = \omega_{\Lambda \beta}.
\end{equation}
Thus the KMS states considered here are in particular  
isotropic in their rest system and invariant under space-time
translations. On the other hand, velocity transformations 
in general change these states, in accordance  with the well 
known fact that the Lorentz group is spontaneously broken in 
thermal equilibrium states \cite{Oj}. Note, however, that the temperature,
fixed by the length of $\beta$, remains unaffected by Poincar\'e
transformations.

So far we have considered only KMS states, where 
the temperature vector $\beta$ is sharply defined. Yet thinking 
of macroscopic ensembles in thermal equilibrium where $\beta$ is 
not precisely known or statistically fluctuating, it is natural 
to consider also mixtures of KMS states. We therefore define 
for any compact subset $\Delt \subset V_+$ corresponding families
$\CD$ of states consisting of all mixtures of KMS states $\omega_\beta$ 
with $\beta \in \Delt$. 

In order to obtain a more explicit description 
of the states in $\CD$ and to avoid 
mathematical subtleties, we assume that the KMS states $\omega_\beta$
are weakly continuous in $\beta$, \ie all functions
\begin{equation} \label{continuity}
\beta \mapsto \omega_\beta(A), \quad A \in {\cal A},
\end{equation}
are continuous. Apart from phase transition points (which 
are excluded here by our uniqueness assumption on $\Cb$),
this property is expected to hold quite generally.
With this input, the states 
$\omega_\Delt \in \CD$ can be represented in the form 
\begin{equation} \label{mixture}
\omega_\Delt(A) = 
\int \! d\rho(\beta) \, \omega_\beta(A), \quad A \in {\cal A},
\end{equation}
where $\rho$ is a positive normalized measure which has support 
in $\Delt$. Note that in the case of degenerate families $\Cb$  
the measure would also have support on the spectrum of all other  
classical observables parameterizing the KMS states. 

The elements of $\CU \doteq \bigcup \, \CD$, where the union extends 
over all compact subsets $\Delt \subset V_+$, will be our reference
states for the characterization of the local thermal observables 
and the interpretation of the local thermal properties of
non-equilibrium systems.  This choice turns out to be 
particularly convenient in the present investigation; yet without  
problems one could also admit reference states $\omega_\Delt$ 
corresponding to non-compact sets $\Delt \subset V_+$.

\section{Local thermal observables}
\setcounter{equation}{0}

Although the algebra ${\cal A}$ contains all local observables
of the underlying theory, it is not well suited 
for the description of the local thermal 
properties of physical states. For, all
elements of ${\cal A}$ have quite fuzzy localization properties due to the 
integration of the point fields with test functions.
So they do not have an interpretation in terms of 
physical properties which can be assigned to space-time points.
Furthermore, ${\cal A}$ contains
an abundance of elements which are of no relevance to  the  
thermodynamic interpretation of physical states. 

For this reason we will proceed first from the algebra ${\cal A}$
to linear spaces ${\cal Q}_{\, x}$ of quadratic forms which can be
interpreted as  observables 
localized at the space-time points $x \in \RR^4$. In a second step
we will then select from each ${\cal Q}_{\, x}$ suitable 
subspaces ${\cal T}_{\, x}$. 
The elements of ${\cal T}_{\, x}$ 
will be regarded as thermal observables at $x$ whose thermal
interpretation is fixed by the thermal reference states  in ${\cal C}$.

\noindent {\bf 3.1 Point fields as idealized local observables} \\
For the step from smeared field operators to point fields, 
let us assume that  in the vacuum sector ${\cal H}$ there hold  
energy bounds of the form  
\begin{equation} \label{ebound} 
\pm \phi(f) \leq \nu (f) \, (1 + H)^k,
\end{equation}
where $f$ is any real test function with compact support, $\nu (\cdot)$
is some Schwartz norm on the space of test functions, $H$ is the 
(positive) Hamiltonian,
and $k > 0$ depends only on the field $\phi$. This condition has a 
clearcut physical interpretation: it says that the observed values
of $\phi(f)$ can become large only  in states of large energy. 
Such energy bounds have been established in many models
\cite{DrFr}. 

As has been shown in \cite{FrHe}, one can proceed from 
(\ref{ebound}) to an improved estimate (for arbitrary 
$l >0$ and corresponding sufficiently large $m > 0$)
\begin{equation}
|| (1 + H)^{-m} \, \phi(f) \, (1 + H)^{-m} || \leq 
c \, \int \! dx \, |(1 - \Delta)^{-l} f(x) |,
\end{equation}
where $|| \cdot ||$ denotes the operator norm in ${\cal H}$
and $\Delta$ is the Laplacian on $\RR^4$. 
It follows from this bound that for sufficiently large $m$
and suitable sequences of test functions $\delta_i, \, i \in \NN$, 
converging to the Dirac $\delta$-function at the space-time point $x$, there 
exists the (norm) limit
\begin{equation}
(1 + H)^{-m} \, \phi(x) \, (1 + H)^{-m} \ \doteq \
\lim_{i \rightarrow \infty} \, (1 + H)^{-m} \, 
\phi(\delta_i) \, (1 + H)^{-m}.
\end{equation}
Thus the field $\phi (x)$ at $x$ is well
defined in the form sense in those states $\omega$ in the vacuum
sector, \cf (\ref{normal}), 
which satisfy $\omega((1+H)^{2m}) < \infty$.
We denote the linear spaces 
of point fields $\phi (x)$, $x \in \RR^4$, for which 
$||(1 + H)^{-m} \phi (x)(1 + H)^{-m} || < \infty $ 
by ${\cal Q}_{\,m,x}, \ m > 0$. These spaces 
are  generically finite dimensional and invariant under
the automorphic action of the stability group of $x$ in 
${\cal P}_+^\uparrow$. Evidently, 
${\cal Q}_{\,m,x} \subset {\cal Q}_{\,m^{\prime},x}$
if $m \leq m^{\prime}$. 

The symmetric elements of ${\cal Q}_{\,m,x}$ may be regarded as 
(idealized) observables at $x$ whose 
expectation values are well-defined in the states characterized 
above. Since the product of fields at a point is meaningless, 
their higher moments are ill-defined, however.
We therefore propose to take as a substitute the 
normal products of these fields which 
can be defined with the help of the Wilson-Zimmermann 
operator-product expansion \cite{WiZi}. 

A novel approach to these notions has recently been 
established by Bostelmann \cite{Bo}. Within this general 
setting one finds that for each $\phi (x) \in {\cal Q}_{\,m,x}$
and spacelike $\zeta \in \RR^4$ the product 
$\phi (x + \zeta) \phi (x - \zeta)$ is well-defined in the 
form sense. Moreover, for any given $q > 0$, there exist a
finite number of distinguished fields $\phi_j (x) \in  {\cal Q}_{\,n,x}$,
$n$ sufficiently large, and analytic 
functions $\zeta \mapsto c_j (\zeta) $, $j = 1, \dots J(q)$,
such that 
\begin{equation} \label{wilson}
|| (1 + H)^{-n} \Big( \phi (x + \zeta) \phi (x - \zeta) 
- \sum_{j=1}^{J(q)} c_j (\zeta) \, \phi_j (x) \Big) (1 + H)^{-n} || 
\leq c \, |\zeta |^q
\end{equation}
for $\zeta$ tending to $0$ in any proper spacelike 
cone. Thus a meaningful substitute
for the ill-defined square of $\phi (x)$ are the subspaces 
${\cal N}(\phi^2)_{\,q,x} \subset {\cal Q}_{\,n,x}$ 
generated by the fields $\phi_j (x)$, $j = 1, \dots J(q)$.
In a similar manner one obtains spaces ${\cal N}(\phi^{\, p})_{\,q,x}$ 
corresponding to higher powers $p$ of $\phi (x)$ \cite{Bo}.
Given $r>0$ and choosing $q$ and $n$ sufficiently large, 
the convergence in (\ref{wilson}) can be made 
sufficiently rapid such that also 
\begin{equation} \label{balance} 
|| (1 + H)^{-n} \Big(\partial_\zeta \, \phi (x + \zeta) \phi (x - \zeta) 
- \sum_{j=1}^{J(q)} \partial_\zeta \, c_j (\zeta) \ \phi_j (x) 
\Big) (1 + H)^{-n} || \leq c \, | \zeta |^{r},
\end{equation}
where $\partial_\zeta$ is any monomial in the partial derivatives
with respect to the components of $\zeta$. Thus the spaces 
${\cal N}(\phi^2)_{\,q,x}$ contain, for large $q$, also 
fields approximating the ``balanced'' derivatives 
$\partial_\zeta \, \phi (x + \zeta) \phi (x - \zeta) $ for 
small $\zeta$.

So far, we have restricted our attention  
to point fields in the vacuum sector. Since we
are interested here primarily in thermal states,  
let us discuss next the circumstances  under which these  
fields can be defined in such states as well.
Here we rely on the fact 
that there exist also local versions of the preceding  
statements in generic cases.
More precisely, in the above estimates the Hamiltonian $H$ can be
replaced by a suitable local Hamiltonian $H_{\cal O}$  with the help of
 \eg the universal localizing maps considered in \cite{BuDoLo}. 
The operators $H_{\cal O}$ induce the time translations of 
the observables in each region ${\cal O} \subset \RR^4$ 
in the sense that the commutators of $H_{\cal O}$ and 
of $H$ with these observables coincide. Roughly speaking, 
the local Hamiltonians $H_{\cal O}$ may be thought of as the 
energy density integrated with suitable smoothed-out characteristic
functions of compact support; so these 
operators can also be used to determine
the local energy content of the states \cite{BuDoLo}.
Thus if a state $\omega$ has locally finite energy
in the sense that $\omega ((1 + H_{\cal O})^{2n}) < \infty$,  
we conclude that the expectation values of all point 
fields $\phi (x) \in {\cal Q}_{\,n,x}$, $x \in {\cal O}$, are well-defined 
in this state.

It is evident that states allowing  locally a thermal
interpretation must have locally finite energy. 
So the upshot of this discussion is the insight that the 
states of interest  can  
be extended to the spaces ${\cal Q}_{\,n,x}$, $n > 0$, 
for $x$ varying in some region ${\cal O}$.
For our thermal reference states 
$\omega_\Delt \in \CU$, this can also be seen more directly 
by noticing that, due to the invariance of these states 
under space-time translations, one has 
\begin{equation} \label{expect}
\omega_\Delt (\phi(x)) = \omega_\Delt (\phi(f))
\end{equation}
for any test function $f$ with $\int \! dx \, f(x) = 1$.   Relation
(\ref{expect}) may be regarded as a formal expression of  
the statement, made in the  introduction, that in thermal equilibrium states
the expectation values of intensive observables can be 
determined in arbitrarily small space-time cells.

\noindent {\bf 3.2 Selection of local thermal observables} \\
As has been indicated above, the algebra ${\cal A}$,
and consequently the spaces ${\cal Q}_{\,n,x}$,
contain many elements which are of no relevance to  the       
thermal interpretation of states at the  
space-time point $x$. Examples are the derivatives
$\partial_x \, \phi (x)$ of point fields which are sensitive to 
the spatio-temporal variations of states and therefore 
vanish in all thermal reference states. So we have to select 
from the spaces ${\cal Q}_{\,n,x}$ suitable subspaces, \cf also
the concluding remarks.
This is accomplished as follows.

For small $n$, the spaces  ${\cal Q}_{\,n,x}$ contain only
multiples of the identity. With increasing   $n$, there will 
appear, at a certain value $n_0 > 0$ in  
the corresponding space ${\cal Q}_{\,n_0,x}$,  
also some point field $\phi_0 (x)$ which we regard as a basic
thermal observable. Next, as has been explained above, 
a meaningful substitute for the square of $\phi_0 (x)$  are 
the normal product spaces ${\cal N}(\phi_0^2)_{\,q,x}$, $q > 0$.
Similarly,  to its higher powers $p \in \NN$ there correspond  the 
spaces ${\cal N}(\phi_0^{\, p})_{\,q,x}$, $q > 0$. Thus the elements 
of all these spaces are to be regarded as thermal observables as well.
They form a proper subspace ${\cal T}_x$ of the space of all point fields,
\begin{equation}
{\cal T}_x \, \doteq \, \sum_{p,q} \, {\cal N}(\phi_0^{\, p})_{\,q,x} \, ,
\end{equation}
where we put
$ {\cal N}(\phi_0^{\, 0})_{\,q,x} \doteq \CC \, 1$ and 
$ {\cal N}(\phi_0^{\, 1})_{\,q,x} \doteq \CC \, \phi_0 (x)$,
$q > 0$.
Assuming for simplicity that the algebra ${\cal A}$
is generated by the averages (\ref{e.smear}) of the 
basic observable $\phi_0$  and its normal products, we are thus led to take 
${\cal T}_x$ as the space of thermal observables at $x$.
Such spaces can also be defined in more general situations
in a systematic manner, but we refrain from giving  
details here. 

The above procedure   
introduces a  natural hierarchy amongst the thermal
observables in ${\cal T}_x$. In order to reveal this fact, 
let us consider any thermal reference state $\omega_\Delt \in \CU$. 
Assuming that the KMS states satisfy the relativistic KMS condition, 
given in  Sec.\ 2, it follows that the two-point functions  
$\zeta \mapsto \omega_\Delt (\phi_0 (x \mp \zeta) \, \phi_0 (x \pm \zeta))$
are the  boundary values of analytic functions in the tips of the  
forward and backward tubes $\RR^4 \pm i \,V_+$, respectively. Furthermore, 
they coincide for spacelike $\zeta$ because of locality.
Thus, by the Edge-of-the-Wedge Theorem,  they  
are analytic for spacelike $\zeta$.  In view of 
the operator-product expansion (\ref{wilson}) and of the invariance 
of $\omega_\Delt \in \CU$ under space-time translations,
these two-point functions are therefore already  
determined by the expectation values of the observables  
$\phi_j (x) \in {\cal N}(\phi_0^2)_{\,q,x}, q > 0$,
in the state $\omega_\Delt \in \CU$. In a similar manner, 
 the spaces ${\cal N}(\phi_0^{\, p})_{\,q,x}, \, q > 0$, 
fix the $p$-point functions for any $p \in \NN$,
unless these functions exhibit essential
singularities at coinciding points.

As the $p$-point functions for large $p$ govern 
the properties of those for smaller $p$
(for the KMS states this is an immediate consequence of the 
cluster property which is, in turn, a consequence of their 
uniqueness),  we conclude 
that the spaces ${\cal N}(\phi_0^{\, p})_{\,q,x}$, $q > 0$,
allow, for increasing $p$, an increasingly finer resolution   
of the properties of the thermal states. As a matter of 
fact, the above argument provides evidence to the effect 
that the set ${\cal T}_x$ of all thermal observables 
at $x$  in general separates the states in $\CU$ and 
is sufficiently big in order to determine 
all properties of the thermal states with arbitrary 
precision. In other words, if one is only interested 
in the gross (macroscopic) properties of these states, 
it suffices to analyze them with the help of some  
subset of thermal observables taken from the spaces
${\cal N}(\phi_0^{\, p})_{\,q,x} \subset {\cal T}_x$ 
for small $p$ and $q$.

\noindent {\bf 3.3 Macroscopic interpretation of local thermal observables}\\
Let us explain next how the elements of ${\cal T}_x$  
provide information about the macroscopic thermal properties 
of the states in $\CU$. In order to simplify this discussion,
we consider only the generic case where ${\cal T}_x$ separates 
these states. 

As we are assuming uniqueness of 
the KMS states, all intensive thermal parameters  
attached to these states can be represented by 
functions $\beta \mapsto \TF(\beta)$ of the temperature vector,
called thermal functions in what follows. 
Hence for mixed thermal states $\omega_\Delt \in \CU$, the mean values 
of the thermal functions $\TF$ are given by 
\begin{equation} \label{macro}
\omega_\Delt (\TF) \doteq \int \! d\rho(\beta) \, \TF(\beta),
\end{equation}
where $\rho$ is the measure appearing in the decomposition
(\ref{mixture}) of $\omega_\Delt$. The notation in (\ref{macro})
indicates that the thermal functions $\TF$ are to be regarded 
as (macroscopic) observables which can be evaluated in all states 
in $\CU$; more precisely, the states can uniquely be extended 
to these functions which appear as limits of 
suitable central sequences of local observables.

It is crucial for our approach that the local observables in 
${\cal T}_x$ provide the same information about the thermal
properties of the states in $\CU$ 
as the macroscopic observables. Namely, one can 
reconstruct with their help   
all relevant thermal functions, in particular the entropy density,
as is shown in the appendix. Moreover, since the thermal observables
${\cal T}_x$ separate the thermal states,  they can be used to determine
the measures $\rho$ in the decomposition (\ref{mixture}), 
which are needed for the evaluation of mean values.

Let us exhibit this fact  
more explicitly. Given any $\phi (x) \in {\cal T}_x$, 
it follows from the transformation formula (\ref{transformation})
that the  function 
\begin{equation} \label{phithermal}
\beta \mapsto \Phi (\beta) \doteq \omega_\beta (\phi(x))
\end{equation}
(which is continuous according to (\ref{continuity}) and 
(\ref{expect}))  does not depend on $x$ and is a Lorentz tensor
(corresponding to  
the tensorial character of $\phi$) built out of   
the temperature vector $\beta$. It may thus be regarded as some
thermal function. We use the fact  that for each $\phi (x)$ the
corresponding function $\Phi$, fixing the thermal interpretation of
$\phi (x)$, is known (by (\ref{phithermal})). As was outlined in the
introduction, this amounts in practice to recording the mean 
values of the local thermal observables in all equilibrium states. 
 In the following we will consistently make use of the 
above notation: lower case Greek letters, such as $\phi$, $\xi$, $\epsilon$, 
denote the microscopic thermal observables whereas the respective upper 
case letters $\Phi$, $\Xi$, $\EE$ denote
the corresponding macroscopic thermal functions. 

Evaluating $\phi (x) \in {\cal T}_x$ 
in an arbitrary state $\omega_\Delt \in \CU$ one obtains 
\begin{equation} 
\omega_\Delt (\phi (x)) = \int \! d\rho(\beta) \, \omega_\beta (\phi (x))
= \int \! d\rho(\beta) \, \Phi (\beta) = \omega_\Delt (\Phi),
\end{equation}
which fixes some (generalized) moment of the {\it a priori}
unknown measure $\rho$. By letting  $\phi (x)$ run through all of 
${\cal T}_x$, the corresponding thermal functions $\Phi$ run
through a dense set of continuous functions on $\Delt \subset V_+$
(since ${\cal T}_x$ separates the states in $\CU$). Then
all moments of the measure are known, hence one can reconstruct
the measure itself from these data, \cf the subsequent section.
At the same time this density property also
implies that the space spanned by the thermal 
functions $\Phi$ can be used to approximate any continuous function 
$\TF$ on compact subsets $\Delt \subset V_+$ with arbitrary precision 
even if $\TF$ is not a member of this space. This is of significance for
quantities such as the entropy density, which may not be expected 
to correspond to any local thermal observable in ${\cal T}_x$. 
In order to determine locally the exact mean values of such thermal  
functions in the states $\CU$, one needs information on the  
expectation values of an infinite number of suitable 
local thermal observables. 

On the other hand, one may expect that basic thermal functions, 
such as the thermal energy density
$\EE$, can in general be determined by some element   
$\epsilon (x) \in {\cal T}_x$. Note, however, that the 
observable $\epsilon (x)$ 
need not coincide with the full stress energy density 
$\theta (x)$ of the underlying theory even though both observables  
have the same expectation values in thermal states, 
\begin{equation}
\omega_\beta (\epsilon (x)) = \EE(\beta) = 
\omega_\beta (\theta (x)),  
\quad \beta \in V_+.
\end{equation}
This peculiarity may be understood if one notices that $\theta (x)$ 
determines not only the local thermal energy of states but also other
forms of energy contained, {\it e.g.}, in the internal stress of 
the system. 
As the energy density of the states $\omega_\beta$ in their rest systems 
is entirely of thermal nature, the observable $\epsilon (x)$ has to 
be sensitive only to this type of energy. This  
point will be exemplified in the model discussed in Sec.\ 5.

\section{Thermal properties of non-equilibrium states}\label{loc-states}
\setcounter{equation}{0}

Having specified the thermal reference states 
and the local thermal observables of the theory, we have now at our 
disposal the necessary tools for the analysis of the local thermal
properties of non-equilibrium states. Throughout
this section we assume without further mentioning that the states 
$\omega$ we are interested in can be extended to 
the local thermal observables. 

\noindent {\bf 4.1 Characterization of locally thermal states}\\
As has been explained in the  introduction, we shall determine the 
thermal properties of arbitrary states by comparing them  
with states in $\CU$ with the help of the local thermal observables. 
Given $x \in \RR^4$ and any subspace ${\cal S}_x \subset {\cal T}_x$, 
we say that a state $\omega$ is ${\cal S}_x$-compatible with 
a thermal interpretation at $x$ (${\cal S}_x$-thermal, for short)
if there exists some $\omega_{\Delt} \in \CU$  such that  
\begin{equation} \label{compare}
\omega (\phi (x)) = \omega_{\Delt} (\phi (x)), \quad \phi (x) 
\in {\cal S}_x.
\end{equation}
Thus $\omega$ looks like a thermal state with respect
to all observables in ${\cal S}_x$. Under these circumstances   
one can consistently define mean values of the  corresponding thermal 
functions
in the state $\omega$ at the space-time point $x$, setting 
\begin{equation} \label{extension}
\omega(\Phi)(x) \doteq  \omega (\phi (x)), \quad \phi (x) 
\in {\cal S}_x,
\end{equation}
where $\Phi$ corresponds to $\phi (x)$, \cf (\ref{phithermal}). 
For if two observables $\phi_1 (x), \phi_2 (x) \in {\cal T}_x$ 
give rise to the same $\Phi$, it follows that  
$\omega_\Delt (\phi_1 (x)) = \omega_\Delt (\Phi) =
\omega_\Delt (\phi_2 (x))$
for all $\omega_\Delt \in \CU$. So, for the given $x$,
the basic relation (\ref{extension}) provides a lift 
of $\omega$ to the subspace of thermal functions $\Phi$
fixed by ${\cal S}_x$. It thereby leads to a local   
thermal interpretation of this state.

The preceding characterization of states admitting
locally a thermal interpretation is physically 
meaningful but difficult to 
use in practice because of the apparent need to
compare these states with all members of the 
family $\CU$. It is therefore gratifying that there 
exists an equivalent characterization which relies 
entirely on the space of 
thermal observables ${\cal T}_x$. 
To see this we introduce on this space a family of seminorms $\tau_\Delt$
for all compact subsets $\Delt \subset V_+$, setting
\begin{equation}
\tau_\Delt (\phi(x)) \doteq \sup_{\beta \in \Delt} \, |\Phi(\beta)|.
\end{equation}
Bearing in mind that the functions $\Phi$ 
are continuous, it is clear that the supremum exists.
The following criterion makes use of this notion.

\nind {\bf Criterion:} Let ${\cal S}_x$ be any subspace of 
${\cal T}_x$ containing the
identity and let $\omega$ be any state on ${\cal A}$. The state 
is ${\cal S}_x$-thermal if and only if there is some
compact subset $\Delt \subset V_+$ such that
\begin{equation} \label{bound}
| \omega (\phi(x)) |  \leq  \tau_\Delt (\phi(x)), \quad 
\phi(x) \in {\cal S}_x.
\end{equation}

The latter condition has a simple physical interpretation: 
the mean values of the local thermal observables 
should not exceed their maximal possible values in the  
thermal states. This constraint can be checked more easily   
 in applications. 

The mathematical 
justification of the criterion relies on standard measure theoretic
arguments: let $\omega$
be ${\cal S}_x$-thermal.
Then there exists some positive, normalized measure $\rho$
with support in $\Delt$, \cf (\ref{mixture}), such that 
\begin{equation}
\omega (\phi(x)) = \omega_\Delt (\phi(x))
= \int \! d\rho(\beta) \, \omega_\beta(\phi(x))
= \int \! d\rho(\beta) \, \Phi (\beta), \quad \phi(x) \in {\cal S}_x.
\end{equation}
Relation (\ref{bound}) then follows by a straightforward
estimate. Conversely, if 
relation (\ref{bound}) holds for $\omega$ and some 
$\Delt$, the linear functional 
$\omega(\Phi)(x) \doteq  \omega (\phi (x))$
on the space of thermal 
functions $\Phi$ corresponding to $\phi (x) \in {\cal S}_x$
is well-defined. For $\Phi = 0$ implies 
$\tau_\Delt (\phi(x)) = \sup_{\beta \in \Delt} \, | \Phi (\beta)| = 0$  
and consequently $\omega(\Phi)(x) = 0$. 
Moreover, as $|\omega(\Phi)(x)| \leq \tau_\Delt (\phi(x)) =
\sup_{\beta \in \Delt} \, |\Phi (\beta)|$, this functional 
can be extended in $\Phi$ by the Hahn-Banach Theorem 
to the space of all continuous functions $\TF$ on $\Delt$ such that 
\begin{equation}
| \omega(\TF)(x) | \leq \sup_{\beta \in \Delt} \, | \TF(\beta)| = ||\TF||_\Delt.
\end{equation}
As the functions $\TF$ with norm $||\, \cdot \,||_\Delt$  form a 
C$^*$-algebra with unit and as  
$\omega(1)(x) = 1$, it follows that 
$\omega(\,\cdot\,)(x)$ defines a positive, normalized functional 
on this algebra \cite{Sa}. But any such functional 
can be represented in the form \cite{Hal}
\begin{equation}
\omega(\TF)(x) = \int \! d\rho(\beta) \, \TF (\beta),
\end{equation}
where $\rho$ is some positive, normalized measure with support
in $\Delt$. Setting 
\begin{equation}
\omega_\Delt (A) \doteq  
\int \! d\rho(\beta) \, \omega_\beta (A), \quad A \in {\cal A}, 
\end{equation}
we have thus found a state $\omega_\Delt \in \CU$ which  
coincides with $\omega$ in the sense of 
condition (\ref{compare}). Hence  $\omega$ 
is ${\cal S}_x$-thermal. 

The thermal states $\omega_\Delt$ describing the local 
thermal properties of $\omega$ are not fixed by condition (\ref{compare})
if the set of thermal observables ${\cal S}_x$ is too small. 
If $\omega$ admits, however, 
a thermal interpretation in the sense of (\ref{compare}) with 
respect to all thermal observables ${\cal T}_x$, then the 
corresponding state $\omega_\Delt \in \CU$ 
(and hence the corresponding measure $\rho$) is in general 
unique by the arguments given in the preceding section.

\noindent {\bf 4.2 Existence of locally thermal non-equilibrium states} \\
For any given finite
dimensional subspace ${\cal S}_x$ of local thermal observables 
and any compact subset $\Delt \subset V_+$, 
there exists an abundance of  
states $\omega$ on ${\cal A}$ which coincide on   
${\cal S}_x$ with some thermal state $\omega_\Delt \in \CU_\Delt$
but are not ${\cal T}_x$-compatible with 
a thermal interpretation. So they correspond to a 
non-equilibrium situation at $x$ which admits, however,  
an interpretation in terms of the subset of thermal 
functions corresponding to ${\cal S}_x$. 

To establish this fact,
let us first assume that $ \tau_\Delt$ defines a norm 
on ${\cal S}_x$, \ie $\tau_\Delt (\phi(x)) = 0$ for 
$\phi(x) \in {\cal S}_x$ implies $\phi(x) = 0$. Picking 
any state $\omega_0$ on ${\cal A}$ and taking into account that on  
a finite dimensional space all linear functionals are 
continuous and all norms are equivalent, we get 
\begin{equation} \label{bound2}
|\omega_0 (\phi (x)) | \leq  c \, \tau_\Delt (\phi (x)), 
\quad \phi (x) \in {\cal S}_x
\end{equation}
for some suitable constant. Thus we can lift 
$\omega_0$ to the space of thermal functions, setting 
$\omega_0 (\Phi) (x) \doteq \omega_0 (\phi (x)) $, $\phi (x) \in {\cal S}_x$. 
Moreover, by the Hahn-Banach Theorem, the latter functional can be 
extended to the space of 
all continuous functions $\TF$ on $\Delt$ such that 
\begin{equation}
| \omega_0 (\TF) (x) | \leq  c \, \sup_{\beta \in \Delt} 
| \TF(\beta) |.
\end{equation}
We may assume that this extension is hermitian, 
replacing $\omega_0 (\TF) (x)$ by the expression
$2^{-1} (\omega_0 (\TF) (x) + \overline{\omega_0 (\TF^*) (x)} )$, 
if necessary. As any linear, hermitian and continuous functional
on the C$^*$-algebra of continuous functions on $\Delt$ can
be represented by a signed measure $\sigma$ with support 
in $\Delt$ \cite{Hal}, we conclude that 
\begin{equation}
\omega_0 (\TF) (x) = \int \! d \sigma(\beta) \, \TF (\beta). 
\end{equation} 
Decomposing $\sigma$ into its positive and negative parts 
$\sigma_\pm$, $\sigma = \sigma_+ - \sigma_-$, and setting 
\begin{equation} \label{perturbation}
\omega (A) \doteq  \big(1 + \sigma_- (\Delt) \big)^{-1} \, 
\big(\omega_0 (A) +  \int \! d \sigma_- (\beta) \,
\omega_\beta (A) \big), \quad A \in {\cal A},
\end{equation}
we have thus exhibited a state $\omega$ on ${\cal A}$ 
such that   
\begin{equation}
\omega (\phi(x)) = 
\omega_\Delt (\phi(x)) \doteq 
\sigma_+(\Delt)^{-1} \, \int \! d \sigma_+ (\beta) \,
\omega_\beta (\phi(x)), \quad \phi (x) \in {\cal S}_x,
\end{equation}
\ie $\omega$ is ${\cal S}_x$-thermal. 
But in general it is not ${\cal T}_x$-thermal
since $\omega_0$ was completely arbitrary.
As is seen from relation (\ref{perturbation}),
the state $\omega$ so constructed 
may be interpreted as a perturbation of a 
thermal background state in $\CU_\Delt$.

Let us turn now to the case where there is some non-trivial subspace 
${\cal S}_{x, \, 0} \subset {\cal S}_x$ which is annihilated 
by $ \tau_\Delt$. If, as expected in the absence of phase transitions,
the thermal functions $\Phi$ are analytic on $V_+$,
the relation $\tau_\Delt ( \phi(x)) = 0$ 
implies $\Phi = 0$ if $\Delt$ has an open interior, \ie
there exist non-trivial relations between the thermal functions 
$\Phi, \, \phi(x) \in {\cal S}_x$. The existence of such 
relations has a simple physical interpretation: it amounts 
to the existence of equations of state (such as the relation between 
energy density and pressure, for example).
 
For the construction of non-equilibrium 
states which are ${\cal S}_x$-thermal, we pick now any 
state $\omega_0$ on ${\cal A}$ which complies with these equations, 
\ie which annihilates the subspace ${\cal S}_{x, \, 0}$.
Such states which are   
not ${\cal T}_x$-thermal  also 
ought to  exist in abundance; for the local validity of 
some equation of state does in general not imply that the 
system is locally in perfect thermal equilibrium
in the sense that all correlations
described by the elements of ${\cal T}_x$ are  
of a thermal nature. We will exemplify this fact in the model discussed
in the subsequent section. 

In view of its proper 
kernel, the functional $\omega_0$ can be projected to the 
(finite dimensional) quotient space ${\cal S}_x / {\cal S}_{x, \, 0}$ 
on which $ \tau_\Delt$ induces a norm. 
Hence, as before, 
$\omega_0$ satisfies the bound 
(\ref{bound2}) and the subsequent construction 
can be carried out. So we conclude that there exist  
non-equilibrium states admitting  
a thermal interpretation for any given 
finite dimensional subspace of thermal observables.

\noindent {\bf 4.3 The degree of thermal stability} \\
Let us discuss next how the states 
admitting locally a thermal interpretation can be ordered
into classes of increasing thermal stability (thermalization). 
Here enters  the hierarchical structure of the   
thermal observables ${\cal T}_x$, exhibited in the preceding
section. Starting with the trivial
subspace ${\cal S}_x = \CC 1$, all
states are ${\cal S}_x$-thermal since only their normalization is tested.
By adding to ${\cal S}_x$ the basic thermal 
observable $\phi_0 (x)$ one imposes already some non-trivial 
constraint. For $\phi_0 (x)$ is an unbounded quadratic form,
hence arbitrary states generically violate condition (\ref{bound}) 
for given $\Delt$.
Next, one adds to ${\cal S}_x$ the elements of 
the normal product spaces ${\cal N}(\phi_0^2)_{\,q,x}$ for 
increasing $q$ which provide information about the correlations 
of the basic observable $\phi_0 (x)$ at neighboring points. 
Using these enlarged spaces ${\cal S}_x$ in the compatibility 
condition (\ref{compare}) one selects subsets of  
states where these correlations are of a thermal nature.  
Note that these constraints resemble the conditions imposed 
on states in the derivation of transport equations.

Proceeding in this manner, one arrives at spaces ${\cal S}_x$ 
containing also elements of ${\cal N}(\phi_0^p)_{\,q,x}$
for higher powers $p$. As has been explained in the preceding
section, the  
resulting compatibility conditions distinguish states  
which exhibit increasingly more refined features of thermal equilibrium 
states and, in this sense, come closer to the idea of a 
genuine equilibrium situation. Thus, for states which are  
${\cal S}_x$-thermal,
the size of ${\cal S}_x$ may be taken as a measure for 
their degree of thermal stability. 

\noindent {\bf 4.4 Determination of specific thermal properties} \\
To judge whether in a given state $\omega$ 
some thermal function $\Phi$ has locally a definite value,
one has to make sure that the 
state is compatible with a thermal \mbox{interpretation}
on some sufficiently large space ${\cal S}_x$. For one has 
to determine not only the mean value of 
$\Phi$ but  also its statistical fluctuations.
If, for example, ${\cal S}_x$ contains observables 
$\phi_1(x)$ and $\phi_2 (x)$
corresponding to $\Phi$ and $\Phi^2$, respectively, then the observable
\begin{equation}
\delta \phi_\kappa (x) \doteq \phi_2 (x) - 2 \kappa \, \phi_1 (x)
+ \kappa^2 \, 1, \quad \kappa \in \RR,
\end{equation}
corresponds to the thermal function $(\Phi - \kappa \, 1)^2$.
It is therefore non-negative in all thermal reference states $\CU$
and vanishes only in those states in which $\Phi$ has the sharp
value $\kappa$. Thus if $\omega$ is ${\cal S}_x$-thermal
and $\omega (\delta \phi_\kappa (x)) = 0$
for some $\kappa$, we may conclude that $\Phi$ has 
the sharp value $\kappa$ at $x$ in this state. It is apparent that 
this conclusion can also be drawn under more general  
conditions. 

For suitable spaces ${\cal S}_x$ one can  also  distinguish in this 
manner states which locally have a sharp temperature
vector $\beta$, \ie which coincide on ${\cal S}_x$ with the
KMS state $\omega_\beta$. The minimal spaces admitting 
such an analysis are finite dimensional in generic cases, 
\cf the discussion in the subsequent section.

If a state $\omega$ coincides on ${\cal S}_x$ 
with some KMS state $\omega_\beta$, all thermal 
functions $\Phi$ corresponding to $\phi (x) \in {\cal S}_x$
have locally definite values in this state. However,  by enlarging
${\cal S}_x$, $\omega$ may cease to have a thermal 
interpretation. Phrased differently, $\omega$  may share only certain   
gross thermal properties with the KMS state $\omega_\beta$
and a more refined analysis would reveal its non-equilibrium 
nature. Thus the method of 
analyzing the local thermal properties of states with the help
of suitable subsets of thermal observables amounts to some
procedure of coarse graining. It leads to the identification 
of an abundance of states having locally certain definite 
thermal properties. Such an approach seems natural if one deals
with non-equilibrium systems. 

\noindent {\bf 4.5 Space-time evolution of thermal properties} \\
The formalism established so far can easily be extended to 
states admitting a thermal interpretation in subregions 
${\cal O} \subset \RR^4$. Yet in order to obtain a sufficiently 
simple description, one has to keep the thermal 
functions fixed   in the respective regions. 
This is accomplished by identifying the 
spaces ${\cal S}_x$, $x \in {\cal O}$, with the help 
of the automorphic action of the translations, setting 
\begin{equation}
{\cal S}_x \doteq \alpha_x ( {\cal S}_0), \quad x \in {\cal O}.
\end{equation}
With this convention understood, we say a state $\omega$ is 
${\cal S}_{\cal O}$-compatible with a thermal interpretation
in ${\cal O}$ if for each $x \in {\cal O}$ there is some
$\omega_\Delt \in \CU$ (in general  depending on $x$) such that
relation (\ref{compare}) holds for the respective ${\cal S}_x$.
The resulting functions 
\begin{equation}
x \mapsto \omega (\Phi) (x) = \omega (\phi(x)), \quad x \in {\cal O},
\end{equation}
then describe the space-time behavior of the mean values of 
the thermal functions $\Phi$. Hence they provide 
the link between
the microscopic dynamics and the evolution of the macroscopic 
thermal properties, \ie the thermodynamics of the states. 
It is an intriguing question under which circumstances this evolution can
also be described in terms of transport equations.

We have thus solved the conceptual problem of identifying
non-equilibrium states admitting locally a thermal 
interpretation, and of describing their specific thermodynamic properties. 
This formalism can now be applied to the analysis of 
non-equilibrium states in concrete models, as will be 
exemplified in Sec.\ 5.
Furthermore,  it is a suitable starting point for 
a general structural analysis of these states.

\noindent {\bf 4.6 Field equations and quasiparticles}  \\     
As a first illustration of the latter point, let us show  
how the microscopic dynamics leads to 
linear differential  equations for the evolution of 
mean values of thermal functions
in states which are sufficiently close to thermal 
equilibrium.  We start from the assumption that 
our basic observable $\phi_0 (x)$ 
satisfies a field equation of the form 
$\square_x \phi_0 (x) = \xi_0 (x)$, where 
$\xi_0 (x)$ is some linear combination 
of elements in the normal product spaces 
${\cal N}(\phi_0^{\,p_0})_{\,q_0,x}$ 
for certain $p_0$ and
$q_0$ (hence it  is a thermal observable).  As  
$\omega_\beta (\xi_0 (x)) = \omega_\beta (\square_x \phi_0 (x)) = 
\square_x \omega_\beta (\phi_0 (x)) = 0$ 
because of the invariance of the KMS-states under 
space-time translations, we conclude that the 
thermal function $\Xi_0$ corresponding to
$\xi_0 (x)$ vanishes. 

Now let ${\cal S}_x$ be any space of thermal 
observables containing the above normal product 
spaces ${\cal N}(\phi_0^{\,p_0})_{\,q_0,x}$,
hence in particular $\phi_0 (x)$ and $\xi_0 (x)$,
and let $\omega$ be any state which 
is ${\cal S}_{\cal O}$-compatible with a 
thermal interpretation in ${\cal O}$.
Then the local mean values of the thermal function 
$\Phi_0$ in this state necessarily satisfy 
the wave equation in ${\cal O}$, as is seen 
from the chain of equalities
\begin{equation}
\square_x \omega(\Phi_0)(x) =  \square_x \omega(\phi_0 (x)) 
= \omega(\square_x \phi_0 (x)) = \omega (\xi_0 (x)) 
= \omega(\Xi_0)(x) =0.
\end{equation}

There exist many other thermal observables for which such 
 behavior of the mean values in suitable states 
can be established. To reveal the underlying simple
mechanism, we proceed from the elementary relation 
\begin{equation} \label{diff}
\begin{split}
& \qquad \quad {\square_x \, \phi_0 (x + \zeta) \phi_0 (x - \zeta) }  \\
& =  - \square_\zeta \, \phi_0 (x + \zeta) \phi_0 (x - \zeta)  
+ 2 \xi_0 (x + \zeta) \phi_0 (x - \zeta)  
+ 2 \phi_0 (x + \zeta) \xi_0 (x - \zeta),
\end{split}
\end{equation}
which holds in the form sense for spacelike $\zeta$.
Taking into account the remark after relation (\ref{balance})
about balanced derivatives, it is clear that 
the expressions on the right hand side of this equality 
can be approximated in the limit of small spacelike $\zeta$ by
elements of the normal product spaces 
${\cal N}(\phi_0^{p})_{\,q,x}$  
for suitable $p$ and $q$. Similarly, 
the expression $\phi_0 (x + \zeta) \phi_0 (x - \zeta)$ 
can be approximated by elements of 
$ {\cal N}(\phi_0^2)_{\,q,x}$. Comparing both sides
of relation (\ref{diff}), one thus finds 
thermal observables $\phi (x)$ in 
${\cal N}(\phi_0^2)_{\,q,x}$
for which $\square_x \, \phi (x) = \xi (x)$ is also a thermal observable. 
As before  one can then show that the local 
mean values of the thermal function $\Phi$ corresponding 
to $\phi (x)$ are solutions of the wave equation in states
which are compatible with a thermal interpretation on 
sufficiently large subspaces of thermal 
observables. 

So the space-time evolution of these thermal functions 
exhibits patterns of a massless particle propagating
through the state $\omega$, provided $\omega$ is locally
sufficiently close to thermal equilibrium. 
Note that this thermal compatibility condition imposes
quite stringent constraints on the state in the case
of interaction, since the spaces ${\cal S}_x$ then have to 
contain thermal observables in ${\cal N}(\phi_0^{\,p})_{\,q,x}$
for $p > 2$. For states slightly violating this condition 
in the sense that $\omega (\xi (x))$ is different from $0$ but
small, the above equations are, however,  
still valid in an approximate sense. This result is 
in accordance with the familiar quasi-particle interpretation 
of perturbations of equilibrium states. It emerges here
as a by-product of our approach to the characterization 
of the local thermal properties of non-equilibrium states.

\section{An instructive example}\label{example}
\setcounter{equation}{0}

In this section we illustrate the preceding
abstract notions and results in the theory of a free massless 
scalar field. We have chosen this particularly simple example  
since it allows the elementary computation of many quantities of 
interest by scaling arguments. After a brief outline of 
the model we will determine the 
structure and physical significance of its local 
thermal observables. We then exhibit interesting examples  
of non-equilibrium states, describing a ``big heat bang'',
for which a definite temperature, thermal energy 
and entropy density can be defined 
at every space-time point in the future cone 
of  some given initial point. 

\noindent {\bf 5.1 The model} \\
The free massless scalar field $\phi_0 (x)$ on $\RR^4$, 
playing  the role of a
basic thermal observable in the present model,  is 
characterized by the field equation and  commutation relation  
\begin{equation} \label{freefield}
\square_x \phi_0 (x) = 0, \quad \ 
[\phi_0 (x_1),\phi_0 (x_2)] = (2 \pi)^{-3} \int \! dp \, e^{-i(x_1-x_2)p}
\, \varepsilon (p_{0}) \delta (p^{2}) \cdot 1.
\end{equation}
It generates a polynomial *-algebra ${\cal A}$, describing
the local observables of the theory. This algebra 
is stable under the actions of the Poincar\'e group 
${\cal P}_+^\uparrow$,
given by $\alpha_{\Lambda,a} (\phi_0 (x)) = \phi_0 (\Lambda x + a)$, 
the dilations $\RR_+$, given by 
$\delta_s (\phi_0 (x)) = s \, \phi_0 (sx)$,
and the gauge group $\ZZ^2$, given by
$\gamma (\phi_0 (x)) = - \phi_0 (x)$.

We restrict attention here to states $\omega$ on ${\cal A}$ which
are gauge invariant, \ie $\omega \circ \gamma = \omega$, so the 
respective $n$-point functions of $\phi_0$ vanish if 
$n$ is odd. The simplest examples of this type are quasifree states.
They are completely determined by their two-point functions 
through the formula
\begin{equation} \label{wick}
\begin{split}
& {\omega}({\phi_0} (x_{1}){\phi_0}(x_{2}) \cdots {\phi_0}(x_{n}))
\\ 
& \doteq
\begin{cases}
\sum_{\mbox{\tiny pairings}} 
{\omega}({\phi_0}(x_{i_1}) \phi_0(x_{i_2})) \cdots 
{\omega} ({\phi_0}(x_{i_{n-1}}){\phi_0}(x_{i_n})) & 
\text{$n$ even,} \\ 
0 & \text{$n$ odd.}
\end{cases}
\end{split}
\end{equation}
Whenever a (generalized) function 
$\omega (\phi_0 (x_1) \phi_0 (x_2))$
is consistent with the constraints imposed by 
(\ref{freefield}) and the positivity condition
$\omega (\phi_0(\overline{f}) \phi_0(f)) \geq 0$ for all
test functions $f$, the functional $\omega$ obtained 
by linear extension from (\ref{wick}) defines a state on ${\cal A}$.
Thus there is  a large supply of simple states from which we will also 
draw the non-equilibrium states considered below.  

\noindent {\bf 5.2 Thermal reference states} \\
It is a well known fact that the algebra ${\cal A}$ has a unique gauge  %KMS-state
invariant KMS state $\omega_\beta$ for                   
each temperature vector $\beta \in V_+$. This state is 
quasifree, so it is determined by its two-point function
given by 
\begin{equation} \label{freekms}
\omega_\beta (\phi_0 (x_1) \phi_0 (x_2)) =
(2 \pi)^{-3} \int \! dp \,
e^{-i(x_1-x_2)p}\varepsilon (p_{0})\delta (p^{2})
\frac{1}{1-e^{-\beta p}}.
\end{equation}
We mention as an aside that these states satisfy the 
relativistic KMS condition and comply with our continuity
assumption (\ref{continuity}). In fact, the respective functions
are analytic on $V_+$, as is easily verified. 

As outlined 
in  Sec.\ 2, the KMS states $\omega_\beta$, $\beta \in V_+$,
fix the convex set $\CU$ of thermal reference states 
which enters into our analysis of the thermal
properties of (gauge invariant) non-equilibrium states.

\noindent {\bf 5.3 Local thermal observables} \\
Let us turn now to the analysis of the 
local thermal observables in this model. We will 
consider  primarily the spaces ${\cal N}(\phi_0^2)_{\,q,x}$,
$q > 0$, which are generated by  
the Wick square of $\phi_0$,  its balanced derivatives and the unit
operator $1$. Introducing the multi-index notation 
$\bmu = (\mu_1, \mu_2, \dots \mu_m)$
and setting 
$\partial_\zeta^\sbmu =
\partial_{\zeta_{\mu_1}}^{ } \cdots \partial_{\zeta_{\mu_m}}^{ }$, 
these balanced derivatives  are defined by
\begin{equation}
\eth_{ }^\sbmu \! : \! \phi_0^2 \! : \! (x)  \doteq 
\lim_{\zeta \rightarrow \,0} \,
\partial_\zeta^\sbmu \big( \phi_0 (x + \zeta)\phi_0 (x - \zeta) 
- \omega_\infty \big(\phi_0 (x + \zeta)\phi_0 (x - \zeta)\big) \, 1 \big),
\end{equation}
where $\omega_\infty$ denotes the vacuum state (which can be 
recovered from (\ref{freekms}) in the limit of large time-like $\beta$).
Note that for odd $m$ the balanced derivatives vanish, 
since $\phi_0 (x + \zeta)\phi_0 (x - \zeta)$ is even 
in $\zeta$ as a 
consequence of locality.

For the determination of the thermal functions corresponding
to these observables one proceeds from the relation 
\begin{equation}
\begin{split}
& {\omega_\beta \big(\phi_0 (x + \zeta)\phi_0 (x - \zeta)\big) - 
\omega_\infty \big(\phi_0 (x + \zeta)\phi_0 (x - \zeta)\big)} \\
& = (2 \pi)^{-3} \int \! \frac{{d{\bp}}}{|{\bp}|} \,
\cos (2 \zeta \underline{p}) \, (e^{\beta \underline{p}} - 1)^{-1}
=  (2 \pi)^{-3} \sum_{n=1}^\infty \int \! \frac{{d{\bp}}}{|{\bp}|} \,
\cos (2 \zeta \underline{p} ) \, e^{- n \beta \underline{p}} \, , 
\end{split}
\end{equation}
where $\underline{p} \doteq (|{\bp}|,{\bp})$. 
The even derivatives with respect to $\zeta$, when applied to the
latter integral, can be replaced 
by $\partial_\beta^\sbmu$, multiplied with appropriate constants.
We omit the remaining simple computations and only state the final result:
\begin{equation} \label{parameters}
\beta \mapsto
\omega_\beta \big(\eth_{ }^\sbmu \! : \! \phi_0^2 \! : \! (x) \big)
= c_m \,\partial_\beta^\sbmu \, (\beta^{2})^{-1},
\end{equation}
where $c_m = 0$ if $m$ is odd and 
$c_m = (-1)^{m/2} (4\pi)^m (m +2)!^{-1} B_{m + 2} $ if 
$m$ is even, $B_n$ being the Bernoulli numbers. It is instructive to
have a closer look at these functions for small $m$. 

\nind (a) The thermal function attached to 
the density $: \! \phi_0^2 \! : \! (x)$ is
$\beta \mapsto (12 \beta^2)^{-1}$,
\ie the square of the temperature (apart from a constant).

\nind (b) The balanced derivatives 
$\epsilon^{\mu\nu}(x) \doteq 
- (1/4) \, \eth_{ }^{\mu \nu} \! : \! \phi_0^2 \! : \! (x)$
give rise to the thermal functions
\begin{equation}
\beta \mapsto \EE^{\mu \nu} (\beta) \doteq 
(\pi^2/90) 
\big( 4 \beta^\mu \beta^\nu - \beta^2 g^{\mu\nu} \big) (\beta^2)^{-3}.
\end{equation}
So they coincide with the expectation values of the 
components of the standard (symmetric and traceless) stress energy
tensor $\theta^{\mu\nu} (x)$ in the states $\omega_\beta$. 
This fact can be understood if one notices that  
this tensor can be represented 
in the form 
\begin{equation} \label{stressenergy}
\theta^{\mu\nu} (x) = \epsilon^{\mu \nu} (x)  
+ (1/12) \,( \partial_{x}^{\mu} \partial_{x}^{\nu} \, 
-  g^{\mu\nu} \, \square_x ) : \! \phi_0^2 \! : \! (x),
\end{equation}
where the second tensor on the right hand side vanishes in all 
equilibrium states because of the derivatives with respect to $x$. 
Since this tensor is the coboundary of the vector-valued 
two-form 
$ (1/12) \, 
(\partial_x^\mu \, g^{\rho \nu} - \partial_x^\rho \,  g^{\mu \nu})  
: \! \phi_0^2 \! : \! (x)$, it follows from Gauss' law that 
it also does not contribute to the total energy of states 
which deviate only locally from an equilibrium situation. 
On the other hand, it is of relevance  
in non-equilibrium states, where it describes the transport
of energy driven by sources which are localized 
at the boundary of the system.

We call $\epsilon^{\mu \nu} (x)$ the thermal energy tensor. 
It is conserved and symmetric, but its trace $\epsilon_\mu^{\, \mu} (x) = 
(1/4) \, \square_x : \! \phi_0^2 \! : \! (x)$ does not vanish
identically. According to the discussion at the end of the 
preceding section, it is, however, 
zero in all states which are sufficiently close to 
equilibrium.

The thermal energy tensor and the Wick square can be used 
to distinguish KMS states in $\CU$ corresponding to
a given temperature vector. To establish this fact we consider, for   
fixed $\varkappa \in V_+$, the thermal observable 
\begin{equation} \label{test}
\delta_\varkappa \epsilon (x) \doteq 
(30/\pi^2)  \varkappa_\mu \varkappa_\nu \epsilon^{\mu \nu} (x) 
- 24  : \! \phi_0^2 \! : \! (x) 
+ (\varkappa^2)^{-1} \, 1
\end{equation}  
with corresponding thermal function
\begin{equation}
\beta \mapsto 
(1/3) 
\big( 4 (\beta \varkappa)^2 - \beta^2 \varkappa^2 \big) (\beta^2)^{-3}  
- 2  (\beta^2)^{-1} + (\varkappa^2)^{-1}. 
\end{equation}
As $(\beta \varkappa )^2 \geq  \beta^2 \varkappa^2$, where  
equality holds if and only if $\beta, \varkappa \in V_+$ are 
parallel, this function is evidently positive on $V_+$
apart from the point $\beta  = \varkappa$, where it vanishes.
Hence if  
$\omega_\Delt (\delta_\varkappa \epsilon (x)) = 0$ 
for some state $\omega_\Delt \in \CU$ one can conclude 
that it must be the KMS state $\omega_\varkappa$ corresponding to 
the temperature vector $\varkappa$. Thus in this model  
a finite number of thermal observables are sufficient  
to decide  whether a state has locally
a definite temperature and a well-defined   rest system.
 
\nind (c) Turning to the higher balanced derivatives, it follows from 
(\ref{parameters}) that all corresponding thermal 
functions are solutions of the wave 
equation since $\square_\beta \, (\beta^2)^{-1} = 0$
on $V_+$.
So the subspace generated by these 
functions does not separate the states in $\CU$. 
Yet it is sufficiently large in order to approximate the entropy 
current $S^\mu$,  
\begin{equation}
\beta \mapsto S^\mu (\beta) \doteq (2 \pi^2/45) \, \beta^\mu
(\beta^2)^{-2},
\end{equation}
on any compact subset $B \subset V_+$. (The determination of the 
entropy current from the microscopic data is outlined in 
the appendix.)

For the proof of this assertion we pick any lightlike vector $l$
and contract the balanced derivative
$\eth_{ }^\sbmu \! : \! \phi_0^2 \! : \! (x)$
with the  tensor $l_\sbmu$. The corresponding 
thermal function is, \cf relation (\ref{parameters}),
\begin{equation}
\beta \mapsto 
c_m \, (l \partial_\beta)^m  \, (\beta^2)^{-1} 
= c_m \, m! \, 2^m \, (l \beta)^m (\beta^2)^{-m-1}
\end{equation}
if $m$ is even; for odd $m$ it is identically zero. As 
$(l \beta) (\beta^2)^{-1}$ is positive
and bounded on  
any compact subset $B \subset V_+$ and the (positive)
square root is holomorphic on the right complex half plane, 
it follows  that $\beta \mapsto (l \beta)
(\beta^2)^{-1}={\big((l\beta)^2 (\beta^2)^{-2} \big)}^{1/2}$  can be 
represented on $B$ as a convergent power series 
involving only even powers $(l \beta)^m (\beta^2)^{-m}$, $m \in 2 \NN_0$.
This proves that $\beta \mapsto l S (\beta)$ can be approximated
on $B$ by linear combinations of the thermal functions 
$\beta \mapsto (l \beta)^m (\beta^2)^{-m-1}$, $m \in 2 \NN_0$.
Since $l$ was arbitrary and the lightlike vectors generate 
a basis of $\RR^4$, we conclude that also the entropy 
current can be approximated on any $B$ with arbitrary precision
by the thermal functions
corresponding to the balanced derivatives. 
Thus any state which admits a thermal 
interpretation on the space generated by 
these derivatives has a well-defined  mean 
entropy density at the space-time point $x$.

\nind {\bf 5.4 Examples of non-equilibrium states} \\
We have already seen in the abstract analysis that there 
exist non-equilibrium states admitting a thermal
interpretation on  finite dimensional subspaces of  
thermal observables. Here we consider
non-equilibrium states which are substantially
closer to a thermal equilibrium situation, namely, these states 
admit a thermal interpretation
on the infinite dimensional spaces ${\cal S}_x$
of thermal observables generated by
${\cal N}(\phi_0^p)_{\,q,x}$, $p=0, \dots 3, \, q > 0$, 
for all $x$ in some future light cone.
Moreover, they have a definite temperature,
thermal energy and entropy density at all of these 
space-time points. 

We are taking  here the simplest states of this 
type in order to illustrate our general method. 
Yet these examples are also  
of some physical interest. They describe 
the spatio-temporal evolution of systems 
which have infinite 
temperature at some space-time point  
(corresponding to a ``big heat bang''). Although
we are dealing  with a massless free
field theory, our results provide some 
idea of the dynamical effects of such singularities
in more realistic theories. For 
the masses of particles and their interaction should 
play a secondary role in the neighborhood
of such singular points.

The non-equilibrium states considered here are quasi-free with 
two-point functions of the form, $\gamma > 0$ being fixed and 
$x_1,x_2 \in V_+$,
\begin{equation} \label{bhb}
\omega_{\mbox{\tiny bhb}} (\phi_0 (x_1) \phi_0 (x_2)) \doteq
(2 \pi)^{-3} \int \! dp \,
e^{-i(x_1-x_2)p}\varepsilon (p_{0})\delta (p^{2})
\frac{1}{1-e^{- \gamma (x_1 + x_2)p}}.
\end{equation}
These functions are 
consistent with the field equation and 
commutation relation in (\ref{freefield}). They are also 
invariant under Lorentz transformations and dilations
but not under translations. Moreover, 
\begin{equation}
\begin{split}
& \omega_{\mbox{\tiny bhb}} \big(\phi_0 (x_1) \phi_0 (x_2) \big) \\[-1pt] 
& = \omega_\infty \big(\phi_0 (x_1)\phi_0 (x_2)\big) 
+ (2 \pi)^{-3} \sum_{n=1}^\infty \int \! \frac{{d{\bp}}}{|{\bp}|} \,
\cos \big( (x_1-x_2) \underline{p} \big) \, e^{- n \gamma (x_1 + x_2) 
\underline{p}} \, ,
\end{split}
\end{equation}
hence, decomposing the cosine according to Euler's formula, 
it follows at once that  each functional $\omega_{\mbox{\tiny bhb}}$
satisfies the positivity condition
$\omega_{\mbox{\tiny bhb}} \big(\phi_0 (\overline{f}) \phi_0 (f)
  \big) \geq 0$ if $\supp f \subset V_+$. By 
standard arguments it can therefore be extended to all of 
${\cal A}$ to a (singular) state. Since we are not 
interested here in this extension we do not dwell upon this point 
any further.  

The computation of the expectation values of the balanced derivatives 
 yields for even $m$ and $x \in V_+$ 
\begin{equation} \label{localkms}
\begin{split}
& \omega_{\mbox{\tiny bhb}} \big(\eth_{ }^\sbmu 
\! : \! \phi_0^2 \! : \! (x) \big) \\
& = (2 \pi)^{-3} \, 
\sum_{n=1}^\infty \int \! \frac{{d{\bp}}}{|{\bp}|} \,  (-1)^{m/2} 2^m 
\underline{p}^\sbmu  \, e^{- n \, 2 \gamma x \, \underline{p}} \ 
= \ \omega_{\beta(x)} \big(\eth_{ }^\sbmu 
\! : \! \phi_0^2 \! : \! (x) \big),
\end{split}
\end{equation}
where $\omega_{\beta(x)}$ is the KMS state 
corresponding to the temperature vector $\beta(x) \doteq 2 \gamma x$.
Thus $\omega_{\mbox{\tiny bhb}}$ is 
${\cal S}_{V_+}$-thermal,
${\cal S}_x$ being the subspaces
of thermal observables generated by 
${\cal N}(\phi_0^p)_{\,q,x}$, $p=0, \dots 3, \ q > 0$.
Note that $\omega_{\mbox{\tiny bhb}} $ 
is compatible with a thermal interpretation on 
the spaces ${\cal N}(\phi_0^p)_{\,q,x}$ for odd $p$ and $q > 0$
since it is, like the KMS states, gauge invariant. 
It has, however, no longer a thermal interpretation
on the spaces ${\cal N}(\phi_0^4)_{\,q,x}$,
$q > 0$, as one verifies by direct computation.
So the higher correlations of the field $\phi_0$
in this state are of a non-thermal nature. 
But the state approaches equilibrium for large time-like
translations $a$
in the sense that for all local observables $A \in {\cal A}$
\begin{equation}
\lim_a \, \omega_{\mbox{\tiny bhb}} \circ \alpha_a (A)
= \omega_\infty (A), 
\end{equation}
\ie $\omega_{\mbox{\tiny bhb}}$ looks asymptotically like the 
vacuum $\omega_\infty$.

Let us turn now to the local thermal interpretation of this 
state. As $\delta_\varkappa \epsilon (x)  \in {\cal S}_x$,
\cf  relation (\ref{test}), and 
$ \omega_{\mbox{\tiny bhb}} \big(\delta_\varkappa \epsilon (x)\big) = 0$
for $\varkappa = \beta(x)$, we conclude that 
$\omega_{\mbox{\tiny bhb}}$ coincides on ${\cal S}_x$
with the KMS state 
$\omega_{\beta(x)}$, $x \in V_+$ (which is also apparent
from relation (\ref{localkms})). Its local temperature
$T(x) = (\beta (x)^2)^{-1/2} = (4 \gamma^2 x^2)^{-1/2}$ 
decreases monotonically
in time-like directions and all Lorentz observers 
moving along some world line $\RR_+ \, e$
register the same temperature at a given time after 
the big heat bang at $0$. 

The thermal energy density 
in the state $\omega_{\mbox{\tiny bhb}}$ 
at the space-time points $x \in V_+$ can be read off from the tensor 
\begin{eqnarray}
%\begin{split}
\omega_{\mbox{\tiny bhb}} (\EE^{\mu \nu}) (x) 
& = & \omega_{\mbox{\tiny bhb}} \big(\epsilon^{\mu \nu} (x) \big)
  =  (\pi^2/90) \big( 4 \beta^\mu(x) \beta^\nu (x) 
- g^{\mu \nu} {\beta (x)}^2 \big) (\beta (x)^2)^{-3} \nonumber \\ 
\qquad \qquad \quad \ \,
& = & (\pi^2/1440 \gamma^4) \big( 4 x^\mu x^\nu - g^{\mu \nu} x^2 \big) 
(x^2)^{-3},
%\end{split}
\end{eqnarray}
describing a flow of massless particles in $V_+$
which is isotropic for the above Lorentz observers.  They also find
that the relation between the thermal energy density and temperature  
is  in accordance with Stefan-Boltzmann's law
for scalar massless particles.

Since $\epsilon_\mu^{\, \mu} (x) = 
(1/4) \, \square_x:\!\phi_0^2\!:\! (x)$ and because 
$\omega_{\mbox{\tiny bhb}} (\EE_\mu^{\, \mu})
(x) = 0$, it follows that 
$\square_x \, \omega_{\mbox{\tiny bhb}} 
(:\!\phi_0^2\!:\!(x)) = 0$,
so the density $:\!\phi_0^2\!:\! (x)$ propagates through 
the state like a massless particle, in accordance with the 
general results in Sec.\ 4. As a matter of fact, 
relation (\ref{localkms}) implies that 
$\square_x \, \omega_{\mbox{\tiny bhb}} \big(\eth_{ }^\sbmu 
\! : \! \phi_0^2 \! : \! (x) \big) = 0$ for all balanced
derivatives. 

It is an interesting fact that the full energy density of the 
state $\omega_{\mbox{\tiny bhb}}$ is larger than its thermal
energy density. Making use of relation (\ref{stressenergy}) one
obtains 
\begin{equation} 
\omega_{\mbox{\tiny bhb}} (\theta^{\mu \nu} (x))
= \big( (\pi^2/1440 \gamma^4) + (1/288 \gamma^2) \big)
\big( 4 x^\mu x^\nu - g^{\mu \nu} x^2 \big) 
(x^2)^{-3}.
\end{equation}
Thus the coboundary term in relation (\ref{stressenergy})
leads to an additional contribution due to the 
transport of energy from the hot boundary 
of the light cone into its interior. Note that for $\gamma = 1$
this term is of the same order of magnitude as the thermal
energy, but it is not visible in the local thermal 
properties of the state. In other words, if one would try 
to determine the energy density of the state from its temperature
by using Stefan-Boltzmann's law, one would 
underestimate its value. This feature 
of an apparently ``missing energy'' may be of significance in 
cosmological models.

As $\omega_{\mbox{\tiny bhb}}$ is compatible with a thermal
interpretation on all spaces 
${\cal N}(\phi_0^2)_{\,q,x}$, $q > 0$, one can consistently
attribute to it the  mean entropy current 
\begin{equation}
\omega_{\mbox{\tiny bhb}} (S^\mu) (x)
= (2 \pi^2/45) \, \beta^\mu (x) (\beta (x)^2)^{-2} =
(\pi^2/180 \gamma^3) \, x^\mu  (x^2)^{-2}
\end{equation}
for $x \in V_+$, \cf
the discussion in the preceding subsection.
Since the model does not describe dissipative effects due to the lack of 
interaction, it does not come unexpectedly that this 
current is conserved, 
\begin{equation} 
\partial_{\mu} \, \omega_{\mbox{\tiny bhb}} (S^\mu) (x)
= 0,
\end{equation}
so there is no entropy production. The entropy current 
decreases monotonically in time-like directions. But the 
total entropy within spacelike sections of $V_+$ 
of the form $|\bx| < v |x_0| $ for fixed $v < 1$ stays constant
for $x_0 > 0$. Thus, from a macroscopic point of view, the 
state $\omega_{\mbox{\tiny bhb}}$ describes 
an equilibrium situation at all points in $V_+$  in spite of the
fact that it is  microscopically out of equilibrium. 

We conclude our discussion of the state $ \omega_{\mbox{\tiny bhb}}$ with 
the remark that one can generate from it other states which are still
${\cal S}_{V_+}$-thermal. 
Namely, for any given translation $a \in V_+$ the state
$\omega_{\mbox{\tiny bhb}} \circ \alpha_a$ has this 
property, too, since $ \alpha_a ({\cal S}_{V_+}) \subset {\cal S}_{V_+}$
and the thermal reference states in $\CU$ are invariant under translations. 
Moreover, since $\CU$ is stable under convex combinations, 
the states $\int \! d\nu(a) \, \omega_{\mbox{\tiny bhb}} \circ
\alpha_a$, where $\nu$ is any positive normalized measure
with compact support in  $V_+$,
are also ${\cal S}_{V_+}$-thermal.
If $\nu$ differs from the Dirac measure, these states have, however, 
locally no longer a definite temperature vector and the corresponding 
expectation values of the  thermal functions exhibit a more 
complex space-time behavior. 

%These results exemplify that our  
%general method for the determination  
%of the thermal properties of non-equilibrium states
%is useful for the analysis of concrete models.

\section{Concluding remarks}\label{conclusion}
\setcounter{equation}{0}

We have established in the present article a novel method for 
the characterization and analysis of non-equilibrium states admitting 
locally a thermal interpretation. The basic idea 
in our approach is to compare 
these states with global equilibrium states by means of local observables. 
The inevitable step of ``coarse graining'' in the passage  
from a microscopic theory to a  macroscopic description
has been accomplished by restricting attention 
to distinguished subspaces of thermal observables, 
in accordance with the basic ideas of statistical 
mechanics having their origin with Boltzmann. 
As the selection of these observables is 
of vital importance in our approach, it seems appropriate  
to comment on this point in somewhat more detail.

At first sight it might seem natural to implement the idea of 
comparing states by defining a suitable distance between them. This can be 
accomplished by introducing a norm $||\, \cdot \,||$  on the
observables ${\cal A}$ (\eg by fixing suitable Schwartz norms on the 
underlying spaces of test functions)
and making use of the inherent net structure 
of the observables, providing to each space-time 
region ${\cal O} \subset \RR^4$ some subalgebra 
${\cal A} ({\cal O}) \subset {\cal A}$ generated by 
observables $A$ whose underlying test functions have 
support in ${\cal O}$, \cf relation (\ref{polynom}).
A physically 
meaningful local norm distance between any two states $\omega_1,\omega_2$
can then be defined, setting 
\begin{equation} \label{norm}
|| \omega_1 - \omega_2 ||_{\, \cal O}
\doteq \sup \, \{ |\omega_1(A) - \omega_2(A)| : A \in {\cal A} ({\cal O}),
\, || A || \leq 1 \},
\end{equation}
which is just the maximal possible difference of the expectation values 
of normalized observables in ${\cal A} ({\cal O})$ in the two
states. 

Making use of this notion, one could try to distinguish 
non-equilibrium 
states $\omega$ admitting a thermal interpretation at space-time point
$x$ by the condition that there is some state 
$\omega_B \in \CU$ and some $q > 0$
such that for each sequence of regions 
${\cal O}_\varepsilon \ni x $ with diameter $\varepsilon$  
\begin{equation} \label{distance}
|| \omega - \omega_B ||_{{\, \cal O}_\varepsilon} 
\leq c \, \varepsilon^q \quad \mbox{for} \quad \varepsilon \rightarrow 0.
\end{equation}
For states $\omega$ coming close to the idea of local 
equilibrium, the exponent $q$ should be large. This condition
turns out to be too stringent, however. Namely if 
$\omega$ has locally finite energy, it 
follows from (\ref{distance}) that  $\omega$ and   
$\omega_B$ coincide on some space of quadratic forms ${\cal Q}_{n(q),x}$,
where $n(q)$ is large for large $q$.
But for sufficiently large $n$ the spaces ${\cal Q}_{n,x}$
contain, together with some observable $\phi (x)$, also its 
derivatives $\partial_x \phi (x)$ \cite{Bo}. Hence  
$\partial_x \, \omega (\phi (x)) = 
\omega (\partial_x \phi (x)) = \omega_B (\partial_x \phi (x)) = 0$
if $\omega$ is, in the above sense, sufficiently 
close to local thermal equilibrium. So, for states 
satisfying the above condition  in some
space-time region,  the expectation values of 
these observables  necessarily 
exhibit a trivial spatio-temporal behavior  in that region.  
In other words, non-equilibrium states which 
are only locally close to equilibrium in general do not comply 
with this strong constraint.

It is noteworthy that one arrives at a similar conclusion
in the C$^*$-algebraic setting of local quantum physics, where 
the forms ${\cal Q}_{n, x}$ appear as dual spaces of the 
germs of states 
of (locally) finite energy \cite{Bo,HaOj}.

We have solved this conceptual problem by distinguishing certain 
subspaces ${\cal T}_x$ of thermal observables, relying on the notion of 
normal product. Intuitively, these observables describe density-like 
quantities which can be attributed to space-time points, 
in contrast to their derivatives, which are sensitive to the 
spatio-temporal changes of states. The spaces  ${\cal T}_x$ have a 
natural hierarchical structure, ranging from observables 
which are sensitive to gross thermal properties up to quantities 
which allow to determine the subtle  
features of higher correlations. 

Non-equilibrium states 
admitting a thermal interpretation can be distinguished
by the condition that they coincide with thermal reference states 
on subspaces of the thermal observables ${\cal T}_x$ and their degree of 
thermal stability can be read off from the size of  
the respective spaces. This characterization is the appropriate
substitute for the apparently too restrictive condition (\ref{distance}). 
 
As has been explained in  Sec.\ 4, the spaces  ${\cal T}_x$
still contain certain
derivatives of observables, but their appearance is of dynamical
origin (it is a consequence of the field equations). Their 
existence leads, on the  one hand, to relations between the thermal
functions, corresponding to equations of state. On the other 
hand, they explain the quasi-particle-like propagation
of densities in states which are sufficiently close to 
thermal equilibrium. 

We have restricted our attention 
to the simplest class of theories, both, with regard 
to the structure of the observables and
the family of thermal reference states.
But our arguments can be extended 
to more complex situations with only little more effort. 
We therefore believe that our approach provides a natural   
setting for the analysis of the thermodynamic properties of 
non-equilibrium states in concrete models as well as in 
the general framework of quantum field theory. It should help to shed new  
light on the complex features of non-equilibrium systems.

A particularly interesting issue is  
the phenomenon of thermalization (approach to 
equilibrium) which is related to the problem 
of the arrow of time. As any local quantum field theory 
is PCT-symmetric and the spaces of thermal 
observables transform covariantly under this 
symmetry, there is no such arrow encoded in the 
algebraic structures used in the present approach to select the states
admitting a thermal interpretation. Yet the states 
can break the PCT symmetry 
in the sense that they are compatible with
a thermal interpretation either in some future or 
in some past light cone, but not in all of Minkowski space
(unless they are in global equilibrium). The state
$\omega_{\mbox{\tiny bhb}}$ considered in Sec.\ 5 
nicely illustrates this point: it does not have a thermal 
interpretation extending beyond $V_+$, and similarly 
the state $\overline{\omega_{\mbox{\tiny bhb}}\circ \vartheta }$, 
obtained from $\omega_{\mbox{\tiny bhb}}$ 
by the action of the anti-automorphism $\vartheta$
implementing the PCT symmetry,  
is thermal only in the past light cone $-V_+$.  
A proof establishing such a one-sidedness of 
thermalization in generic cases would be 
an important step towards the understanding 
of the microscopic origin of the arrow of
time. We hope to return to this problem elsewhere. 

\begin{appendix}
\section{Thermostatics in $\CU$}\label{thermo-K}
\setcounter{equation}{0}

In this appendix we determine the thermostatic
properties of the thermal reference states. As in the 
main text, we assume that for each temperature vector
$\beta \in V_+$ there is a unique KMS state $\omega_\beta$.
By application  of the fundamental laws of thermodynamics,
the thermal functions can be determined if the 
corresponding expectation 
values of the stress energy tensor $\theta^{\mu \nu} (x)$ or 
of the thermal energy tensor 
$\epsilon^{\mu \nu} (x) \in {\cal T}_x$, respectively, 
are known \cite{Di}. We will discuss in   
which sense these functions retain their 
interpretation in thermal reference states which are 
no proper equilibrium states. 

In view of the tensor character of $\theta^{\mu \nu} (x)$ and
relation (\ref{transformation}), the corresponding 
thermal function $E^{\mu \nu}$ has the form
\begin{equation} \label{thermal-energy}
\beta \mapsto E^{\mu \nu} (\beta)
= \omega_\beta (\theta^{\mu \nu} (x)) = 
Q(\beta^2) \, e^\mu e^\nu - P(\beta^2) \, g^{\mu \nu},
\end{equation}
where $e = (\beta^2)^{-1/2} \, \beta$ specifies the
rest system of the state and 
$Q, P$   depend on the underlying
theory. As we are dealing with an 
equilibrium situation, the function $P$ can be interpreted as 
pressure and 
$Q(\beta^2) = - 2 \beta^2 P^{\prime} (\beta^2) $ 
\cite[Ch.\ 4]{Di}. 
Because of relativistic covariance, the entropy current $ S^\mu$
is  of the form 
\begin{equation} \label{thermal-entropy}
\beta \mapsto S^\mu (\beta) = S(\beta^2) \, e^\mu,
\end{equation}
where $S$ is fixed by the Gibbs relation in the rest system 
\cite[Ch.\ 4]{Di},
\begin{equation} \label{entropy2}
S(\beta^2) = (\beta^2)^{1/2} Q(\beta^2).
\end{equation}
The free energy density $F^{\mu \nu}$ is given by the tensor  
\begin{equation}
\beta \mapsto F^{\mu \nu} (\beta) = E^{\mu \nu} (\beta)
- (1/2) (\beta^2)^{-1/2} (e^\mu S^\nu(\beta) + e^\nu
S^\mu(\beta)).
\end{equation}
Inserting the preceding expressions for the entropy and
energy, one finds that it coincides in all 
Lorentz frames with the negative pressure,
\begin{equation}
F^{\mu \nu} (\beta) = - P(\beta^2) \, g^{\mu \nu}.
\end{equation}

Turning to the interpretation of these functions in the 
thermal reference states $\omega_B = \int \! d\rho(\beta) \,
\omega_\beta$, 
there appears no problem with the energy density
which can be determined by the local observable $\theta^{\mu \nu} (x)$.
Hence its expectation values are
\begin{equation}
\omega_B (E^{\mu \nu}) = \int \! d\rho(\beta) \, E^{\mu \nu}
(\beta).
\end{equation}
It is less obvious, however, that the expectation values of $S^\mu$,
\begin{equation} \label{entropyvector}
\omega_B (S^\mu) = \int \! d\rho(\beta) \, S^\mu 
(\beta),
\end{equation}
can be interpreted as entropy currents of the non-equilibrium
states $\omega_B$. In order to justify this interpretation,
we make the following physically meaningful assumptions: the
entropy density $S$ is a concave function of the  
energy density $E = Q - P$, and the pressure $P$ as well as  
$E - P$ increase monotonically with the temperature $T = (\beta^2)^{-1/2}$.
The latter condition means that the negative free energy 
(the isothermal work which can be extracted from 
equilibrium states in their rest systems) 
increases with increasing
temperature, yet less rapidly than the total energy.

We will show that these assumptions entail a version of the second 
law based on (\ref{entropyvector}). 
To this end let us determine in any given Lorentz frame  
$e^\prime \in V_+$, $e^{\prime \, 2} = 1$, the states 
$\omega_B \in {\cal C}$ with fixed energy density
$e^{\prime}_{\mu} e^\prime_{\nu} \, \omega_B (E^{\mu \nu}) = E^\prime$
which maximize the entropy functional 
$\omega_B \mapsto e^\prime_\mu \omega_B (S^\mu)$.
Considering first the KMS states $\omega_\beta$ and making the
substitution $\beta = (\beta^2)^{1/2} e \longrightarrow (E, e)$, we have 
\begin{eqnarray}
& e^\prime_\mu e^\prime_\nu \, \omega_\beta (E^{\mu \nu}) 
= \big( (E + P(E))\, (e^\prime e)^2 - P(E) \big) 
= E^\prime &  \label{g1a} \\
& e^\prime_\mu \, \omega_\beta (S^\mu) =  
S(E) \, (e^\prime e). & \label{g2a}
\end{eqnarray}
Inserting  $(e^\prime e)$ from relation (\ref{g1a}) into (\ref{g2a}),
we get
\begin{equation} \label{g3a}
e^\prime_\mu \, \omega_\beta (S^\mu) =  
S(E) \big((E^\prime + P(E))/ (E + P(E)) \big)^{1/2}.
\end{equation}
The logarithmic derivative of the right hand side 
of this equality with respect to $E$ satisfies 
\begin{equation}
\begin{split}
& (dS/dE)/S + (dP/dE)/2 \big(E^\prime + P \big)
- \big(1 + dP/dE \big)/2 \big(E + P \big) \geq \\
& 1/ST - \big(1 + dP/dE \big)/2 \big(E + P \big) 
=  \big(1 - dP/dE \big)/2 \big(E + P \big) \geq 0,
\end{split}
\end{equation}
where we made use of (\ref{entropy2}) and of the preceding assumptions,
implying $0 \leq dP/dE \leq 1$.
Thus the expression (\ref{g3a}) attains its maximum
for the maximal possible energy $E = E^\prime$ 
(corresponding to $e = e^\prime$) compatible with 
(\ref{g1a}), \ie
\begin{equation} \label{g4a}
e^\prime_\mu \, \omega_\beta (S^\mu) = S(E) \, (e^\prime e) \leq S(E^\prime).
\end{equation}
Turning to the general case, let 
$\omega_B = \int \! d\rho (\beta) \, \omega_\beta $
be any state with
\begin{equation}  \label{g1} 
e^\prime_\mu e^\prime_\nu \, \omega_B (E^{\mu \nu}) = \int \! d\sigma (E,e) \,
\big( (E + P(E))\, (e^\prime e)^2 - P(E) \big) = E^\prime,    
\end{equation}
where $\sigma$ is the normalized measure obtained
from $\rho$ by the substitution $\beta \rightarrow (E,e)$.
In view of the preceding results and the concavity of $S$ we 
obtain
\begin{eqnarray}  \label{g2}
e^\prime_\mu \, \omega_B (S^\mu) & = & \int \! d\sigma (E,e) \, 
S(E) \, (e^\prime e)
\leq  \int \! d\sigma (E,e) \, S\big( (E + P(E))\, (e^\prime e)^2 - P(E)
\big) \nonumber \\ 
& \leq & S \Big( \int \! d\sigma (E,e) \, 
\big( (E + P(E))\, (e^\prime e)^2 - P(E) \big) \Big) = S(E^\prime),
\end{eqnarray}
and  equality holds  in (\ref{g2})
if and only if the measure $\sigma$ is concentrated at the 
point $E^\prime,e^\prime$. Thus $\omega_{\beta^\prime}$,
where $\beta^\prime = (\beta^{\prime \, 2})^{1/2} e^\prime$ and 
$\beta^{\prime \, 2}$ is
fixed by $E^\prime$, is the unique state in ${\cal C}$ which maximizes 
the entropy functional $\omega_B \mapsto e^\prime_\mu \, \omega_B (S^\mu)$
under the given conditions. 
As this functional 
is also additive, 
the interpretation of (\ref{entropyvector}) as entropy current
is justified.

Similarly, one can define the mean free energy density 
of the states in $\CU$,
\begin{equation}
\omega_B (F^{\mu \nu}) = \int \! d\rho(\beta) \, F^{\mu \nu} (\beta)
= - \int \! d\rho(\beta) \, P(\beta^2) \ g^{\mu \nu}.
\end{equation}
It thus coincides with the negative mean pressure in all Lorentz
frames and has a sharp value for all states $\omega_B$ with a definite
temperature $T$, corresponding to measures $\rho$ which are concentrated on the
manifold $\beta^2 = T^{-2}$. Hence 
the set of states minimizing the free energy functional 
$\omega_B \mapsto e^\prime_\mu e^\prime_\nu \, \omega_B (F^{\mu \nu})$
for given temperature is degenerate.

We conclude this appendix by illustrating the preceding notions for  
the case of a dilation invariant theory, where the stress 
energy tensor $\theta^{\mu \nu} (x)$
is traceless, $\theta_\mu^{\, \mu} (x) = 0$, 
and transforms canonically under the dilations 
according to 
$\delta_s (\theta^{\mu \nu} (x)) = s^4 \, \theta^{\mu \nu} (sx)$, $s
\in \RR_+$.
A special example is the model discussed in Sec.\ 5.
It follows from the definition of KMS states and their uniqueness
that $\omega_\beta \circ \delta_s = \omega_{s^{-1} \beta}$.
Plugging these pieces of information into equation (\ref{thermal-energy}),
we obtain 
\begin{equation}
E^{\mu \nu} (\beta) = C \, (4 \beta^\mu \beta^\nu - 
\beta^2 g^{\mu \nu})(\beta^2)^{-3},
\end{equation}
where $C$ is a constant depending on the underlying theory. Thus
the temperature dependence of the 
energy density in the rest system is given by 
$E(\beta) = 3 \, C (\beta^2)^{-2}$, 
in accordance with Stefan-Boltzmann's law, and the pressure satisfies 
\begin{equation}
P(\beta) = C \, (\beta^2)^{-2}.
\end{equation} 
So $E = 3 P$, which also 
follows directly from the fact that the stress energy tensor
is traceless. In particular, $0 \leq dP/dE \leq 1$,
as was anticipated in the preceding analysis. Finally,
the entropy current is given by 
\begin{equation}
S^\mu (\beta) = 4 \, C \, \beta^\mu (\beta^2)^{-2},
\end{equation}
so the pertinent thermal functions are completely fixed 
apart from the value of the constant $C$.
\end{appendix}

\vspace*{5mm}
\noindent {\Large \bf Acknowledgements}\\[2mm]
Our collaboration was made possible by financial support
from Akademie der Wissenschaften zu G\"ottingen,
Japanese Society for the Promotion of Science, 
RIMS Kyoto University, and Universit\"at G\"ottingen,
which we gratefully acknowledge. We are also grateful
for hospitality shown to us by these institutions
at several occasions. 

\end{document}